\documentclass[aps,pra,reprint,superscriptaddress,floatfix,10pt]{revtex4-2}

\usepackage{amsmath}
\usepackage{amsfonts}
\usepackage{amssymb}
\usepackage{graphicx}
\usepackage{color}
\usepackage{braket}
\usepackage{hyperref}
\usepackage{todonotes}
\usepackage{shortvrb}
\usepackage{cprotect}
\usepackage{overpic}
\usepackage[capitalise]{cleveref}

\usepackage{physics}
\usepackage{mathtools}

\definecolor{darkblue}{rgb}{0.00,0.00,0.55}
\definecolor{black}{rgb}{0.00,0.00,0.00}
\hypersetup{
    linkcolor = darkblue,
    anchorcolor = darkblue,
    citecolor = darkblue,
    filecolor = darkblue,
    urlcolor = darkblue,
    colorlinks  = true,
}
\definecolor{brightcerulean}{rgb}{0.11, 0.67, 0.84}
\DeclareMathAlphabet\mathbfcal{OMS}{cmsy}{b}{n}

\newcounter{fig}

\newcommand{\figwidth}{8.6cm}
\begin{document}

\title{Two-Component 3D Atomic Bose-Einstein Condensates Support Complex Stable Patterns}

\author{N.~Boull\'e}
\email{nb690@cam.ac.uk}
\affiliation{Isaac Newton Institute for Mathematical Sciences, University of Cambridge, Cambridge CB3 0EH, United Kingdom}
\author{I.~Newell}
\email{newell@maths.ox.ac.uk}
\affiliation{Mathematical Institute, University of Oxford, Oxford OX2 6GG, United Kingdom}
\author{P.~E.~Farrell}
\email{patrick.farrell@maths.ox.ac.uk}
\affiliation{Mathematical Institute, University of Oxford, Oxford OX2 6GG, United Kingdom}
\author{P.~G.~Kevrekidis}
\email{kevrekid@math.umass.edu}
\affiliation{Department of Mathematics and Statistics, University of Massachusetts
  Amherst, Amherst, MA 01003-4515, USA}

\date{\today}

\begin{abstract}
We report the computational discovery of complex, topologically charged, and spectrally stable states in three-dimensional multi-component nonlinear wave systems of nonlinear Schr{\"o}dinger type. While our computations relate to two-component atomic Bose-Einstein condensates in parabolic traps, our methods can be broadly applied to high-dimensional, nonlinear systems of partial differential equations. The combination of the so-called deflation technique with a careful selection of initial guesses enables the computation of an unprecedented breadth of patterns, including ones combining vortex lines, rings, stars, and ``vortex labyrinths''. Despite their complexity, they may be dynamically robust and amenable to experimental observation, as confirmed by Bogolyubov-de Gennes spectral analysis and numerical evolution simulations.
\end{abstract}


\maketitle


\section{Introduction}
The realm of nonlinear Schr{\"o}dinger models has been one of the principal pillars for the study of nonlinear wave phenomena in dispersive systems~\cite{sulem,ablowitz,ablowitz1,siambook}. The relevant applications span a wide range of fields including nonlinear optical systems~\cite{Kivshar2003}, water waves and plasmas~\cite{infeld}, as well as importantly over the last two decades, the atomic physics setting of Bose--Einstein condensates (BECs)~\cite{pethick,stringari}. In the latter setting, nonlinear structures in the form of bright~\cite{tomio} and dark~\cite{djf} solitary waves, but also importantly topologically charged patterns in the forms of vortices~\cite{fetter1,fetter2} in two dimensions (2D), as well as vortex rings, lines or knots, among others~\cite{komineas,irvine,dsh2} in three dimensions (3D) have played a central role not only in theoretical and computational, but also  in experimental studies. Indeed, these have been connected to notions such as persistent currents~\cite{PhysRevLett.99.260401}, turbulence and associated cascades~\cite{TSATSOS20161}, emulations of an expanding universe in the lab~\cite{PhysRevX.8.021021},  as well as Hawking radiation from analogue black holes~\cite{blackhole}.

While the majority of the relevant contributions focused on single-component systems, such as single atomic species BECs, gradually this situation is changing. Over the past few years, it has been realized that coupled systems can be exploited to manipulate the spin degree of freedom~\cite{RevModPhys.85.1191} in order to produce a wide variety of topological and non-topological, ground and excited state coherent structures, both in one dimension (1D)~\cite{KEVREKIDIS2016140} and higher dimensions~\cite{KAWAGUCHI2012253}. However, developing numerical methods to compute nonlinear waveforms in high dimensions, such as 3D, poses considerable challenges, even in single-component settings~\cite{mateo2015stability,boulle2020deflation}. This is even harder in multi-component systems, where only a few groups have attempted to provide a description of stability of singular and non-singular patterns featuring vortices, monopoles, and the so-called Alice rings~\cite{PhysRevA.86.013613,PhysRevA.93.033633,https://doi.org/10.48550/arxiv.2112.12723,PhysRevLett.91.190402}.

In this work we report on a computational investigation of the solutions of two-component atomic BECs in a three-dimensional parabolic trap. The solutions are discovered with a numerical technique called deflation~\cite{farrell2015deflation}, which has been successfully applied to lower-dimensional or single-component systems~\cite{charalampidis2018computing,charalampidis2019bifurcation,boulle2020deflation}, but has not been extended, to the best of our knowledge, to multi-component 3D problems. We complement this with a Bogolyubov-de Gennes (BdG) stability analysis and transient numerical simulations. Surprisingly, and contrary to what was found to be the case in the single-component setting~\cite{charalampidis2018computing,boulle2020deflation}, the multi-component system allows for the \emph{dynamically robust} existence of unexpected and highly complex vortical states, including ones featuring labyrinthine patterns. This suggests the potential observability of the obtained states.

\section{Setup \& Method}
A mixture of two bosonic components of the same atom species can be described, at the mean-field level, by a system of two coupled Gross--Pitaevskii (GP) equations~\cite{pethick,stringari}. We refer to the two components as $-$ and $+$, and describe their distributions by the wave-functions $\Phi_{\pm}:D \times  \mathbb{R}^+ \to \mathbb{C}$ with spatial domain $D=[-6,6]^3 \subset \mathbb{R}^3$. The model can be written in its non-dimensional form as~\cite{siambook}:
\begin{equation}\label{eq_time_eq}
\begin{aligned}
i\pdv{\Phi_-}{t} \!=\!& -\frac{1}{2}\nabla^2{\Phi_-} \!+\! (g_{11}|\Phi_-|^2 \!+\! g_{12}|\Phi_+|^2)\Phi_- \!+\! V(\mathbf{r})\Phi_-,\\
i\pdv{\Phi_+}{t} \!=\!& -\frac{1}{2}\nabla^2{\Phi_+} \!+\! (g_{12}|\Phi_-|^2 \!+\! g_{22}|\Phi_+|^2)\Phi_+
\!+\! V(\mathbf{r})\Phi_+.
\end{aligned}
\end{equation}
with homogeneous Dirichlet boundary conditions on the boundary of the domain $D$. The symmetric $2\times 2$ coefficient matrix $(g_{ij})_{1\leq i,j\leq 2}$ characterizes the interactions between the two components. Focusing on the case of the hyperfine states of $^{87}$Rb, we use the parameter values $g_{11} = 100.4/98.006$, $g_{12} = 1$, and $g_{22} = 95.44/98.006$, proposed in~\cite[Table~2]{egorov2013measurement}.
The function $V(\mathbf{r})=\frac{1}{2}\Omega^2|\mathbf{r}|^2$ is a parabolic external confining spherical potential with strength $\Omega=1$, where $|\mathbf{r}|^2 = x^2+y^2+z^2$. We compute stationary solutions to the coupled GP equations by assuming the standing wave ansatz $ \Phi_{\pm}(\mathbf{r}, t) = \phi_{\pm}(\mathbf{r})e^{-i\mu_{\pm}t}$, and solving the following system of coupled equations:
\begin{align}
-\frac{1}{2}\nabla^2{\phi_-} \!+\! (g_{11}|\phi_-|^2 \!+\! g_{12}|\phi_+|^2)\phi_- \!+\! V(\mathbf{r})\phi_- \!-\!\mu_-\phi_-&\!=\!0, \nonumber \\
-\frac{1}{2}\nabla^2{\phi_+} \!+\! (g_{12}|\phi_-|^2 \!+\! g_{22}|\phi_+|^2)\phi_+\!+\! V(\mathbf{r})\phi_+\!-\!\mu_+\phi_+&\!=\!0. \label{eq_GP}
\end{align}

We discretize the real and imaginary components of $\phi_-$ and $\phi_+$ using cubic Lagrange finite elements defined on a hexahedral mesh, with ten cells along each axis. The use of a hexahedral mesh is desirable because the discretized problem inherits some of the reflective symmetries of the infinite-dimensional problem. \cref{eq_GP} is solved using the Firedrake finite element library~\cite{rathgeber2016firedrake} by combining Newton's method with the MUMPS LU solver~\cite{amestoy2001} via PETSc~\cite{petsc_user_ref} to solve the resulting linear equations.

\subsection{Deflation of solutions}
We compute multiple solutions to \cref{eq_GP} at parameters $(\mu_-,\mu_+)=(4,5)$ using \emph{deflation}~\cite{farrell2015deflation}. Deflation allows us to identify new stationary solutions by modifying the nonlinear problem solved to prevent the discovery of known solutions by Newton's method. If $F(\phi_-,\phi_+)$ denotes the coupled NLS operator associated with \cref{eq_GP} and $(\phi_{1-},\phi_{1+})$ is a steady-state already obtained, then we construct and solve a new problem $G(\phi_-,\phi_+) = \mathcal{M}_1(\phi_-,\phi_+) F(\phi_-,\phi_+)$. The deflation operator $\mathcal{M}_1$ we employ is
\[\mathcal{M}_1=\left(\frac{1}{\left\||\phi_-|^2 \!-\! |\phi_{1-}|^2\right\|^2}\!+\!1\right)\!\!\left(\frac{1}{\left\||\phi_+|^2\!-\! |\phi_{1+}|^2\right\|^2}\!+\!1\right)\!,\]
where $\|\cdot\|$ is the $H^1(D)$-norm. This operator prevents the convergence of Newton's method applied to $G$ to the previous solution $(\phi_{1-},\phi_{1+})$ (or its multiple by $e^{i\theta}$ for any $\theta \in \mathbb{R}$) since $\mathcal{M}_1(\phi_-,\phi_+)\to \infty$ as $(\phi_{-},\phi_{+})$ approaches $e^{i \theta} (\phi_{1-},\phi_{1+})$. The deflation procedure can be iterated to deflate an arbitrary number of known solutions $\{(\phi_{i-},\phi_{i_+})\}_{i=1}^n$ by constructing a problem $G = \mathcal{M}_n\cdots \mathcal{M}_1 F$. We remark that this operator differs significantly from the one previously used in the single component setting~\cite{boulle2020deflation}. In this work, we employ a factorization of the operator into a product of two deflation operators associated with the first and second components, $\phi_{1-}$ and $\phi_{1+}$, to reduce the number of uninteresting solutions obtained by deflating states of the form $(\phi_{1-},0)$ and $(0,\phi_{1+})$. Solutions of this form that we exclude can be obtained by solving the one-component problem, whereas we are interested in finding steady-states with non-zero coupling between the components.

Given the computational difficulty of the problem, a key challenge is to initialize our search with suitable initial guesses to discover a large number of solutions with complex patterns. Contrary to the single-component setting~\cite{boulle2020deflation}, we found that exploiting linear low-density limits of the system~\eqref{eq_GP} is not an efficient strategy as it requires a large number of Newton iterations to eventually converge and results in simple steady-states. To achieve our goal of discovering complex but experimentally observable solutions (see the discussion of their stability below), we provided Newton's method with a large number of initial guesses of the form $(\phi_-,\phi_+)=(\phi_{7/2},\phi_{9/2})$ by combining solutions to the nonlinear one-component equation. Here $\phi_{7/2}$ and $\phi_{9/2}$ are steady-states of the one-component problem emanating from the third and fourth excited states at chemical potential $\mu=7/2$ and $\mu=9/2$ previously discovered in~\cite{boulle2020deflation}.

Once a steady state has been discovered by deflation at parameters $(\mu_-,\mu_+)=(4,5)$, we continue it to the linear limit at $(\mu_{0-},\mu_{0+})$ by performing a linear interpolation in both components. As $\mu_+$ (the chemical potential of the first component) decreases from $\mu_{\text{init}+}$ to $\mu_{0+}$, we want $\mu_-$ (the chemical potential of the second component) to vary from $\mu_{\text{init}-}$ to $\mu_{0-}$. This yields the following linear interpolation equation for $\mu_-$:
\begin{equation} \label{eq_mu_-}
\mu_- = \mu_{\text{init}-}\frac{\mu_+ - \mu_{0+}}{\mu_{\text{init}+} -\mu_{0+}} + \mu_{0-}\frac{\mu_{\text{init}+}-\mu_+}{\mu_{\text{init}+} -\mu_{0+}}.
\end{equation}
After identifying the chemical potential parameters $\mu_{0-}$ and $\mu_{0+}$ at which the two components emerge, we then discretize the interval $[\mu_{0+}, \mu_{\text{init}+}]$ with regular step-size $\Delta \mu = 10^{-2}$, where $\mu_{\text{init}+}=5$. A steady state is continued to the linear limit as $\mu_+$ goes from $\mu_{\text{init}+}\to \mu_{0+}$ (and similarly for $\mu_-$ using \cref{eq_mu_-}) by solving the NLS system using the solution at the previous step in the chemical potential as initial guess. We then display the continuation of the state $(\phi_-, \phi_+)$ by reporting the atomic number of each component:
\[N_- = \int_\Omega |\phi_-|^2\, \textup{d} x, \qquad N_+ = \int_\Omega |\phi_+|^2\, \textup{d} x,\]
as a function of the parameter $\mu_+$. As the state is continued
towards the low-density limit at $(\mu_{0-},\mu_{0+})$, the atomic numbers $N_-$ and $N_+$ converge to zero.

\subsection{Stability analysis and transient simulations}
We now provide details about the stability computations of the discovered solutions to the time-dependent NLS system \eqref{eq_time_eq}. Once a steady-state $\phi_{\pm}^{0}(\mathbf{r})$ to \cref{eq_time_eq} has been identified by deflation, we perform a Bogolyubov–de Gennes (BdG) spectral stability analysis~\cite{pethick,stringari,siambook} by using the following perturbation ansatz~\cite{charalampidis2019bifurcation},
\begin{equation} \label{lin_ansatz}
\begin{aligned}
\widetilde{\Phi}_{-}(\mathbf{r},t)&=e^{-i\mu _{-}t}\left[\phi _{-}^{0}+\varepsilon
\left( a(\mathbf{r})e^{i\omega t}+b^{\ast }(\mathbf{r})e^{-i\omega ^{\ast }t}\right) \right],\\
\widetilde{\Phi}_{+}(\mathbf{r},t)&=e^{-i\mu _{+}t}\left[ \phi _{+}^{0}+\varepsilon
\left( c(\mathbf{r})e^{i\omega t}+d^{\ast }(\mathbf{r})e^{-i\omega ^{\ast }t}\right) \right],
\end{aligned}
\end{equation}
where $\omega\in\mathbb{C}$ is the eigenfrequency, $\varepsilon\ll 1$ is a small parameter, and $\ast$ denotes the complex conjugate. Inserting this equation into \cref{eq_time_eq} yields an eigenvalue problem at order $\mathcal{O}(\epsilon)$, which we write in matrix form as
\begin{equation}  \label{eig_prob}
\begin{pmatrix}
A_{11} & A_{12} & A_{13} & A_{14} \\
-A_{12}^{\ast } & -A_{11} & -A_{14}^{\ast } & -A_{13}^{\ast } \\
A_{13}^{\ast } & A_{14} & A_{33} & A_{34} \\
-A_{14}^{\ast } & -A_{13} & -A_{34}^{\ast } & -A_{33}%
\end{pmatrix}
\begin{pmatrix}
a \\
b \\
c \\
d%
\end{pmatrix}%
=
\rho
\begin{pmatrix}
a \\
b \\
c \\
d%
\end{pmatrix},
\end{equation}
with eigenvalue $\rho =-\omega $ and eigenvector $\mathcal{V}=\left[a\,b \,c \,d\right]^{\top}$. The matrix elements in \cref{eig_prob} are given by
\begin{align*}
&A_{11} =-\frac{1}{2}\nabla^{2}+\left(2g_{11}|\phi _{-}^{0}|^{2}+g_{12}|\phi _{+}^{0}|^{2}\right) +V(\mathbf{r})-\mu _{-},\\
&A_{33} =-\frac{1}{2}\nabla^{2}+\left(g_{12}|\phi _{-}^{0}|^{2}+2g_{22}|\phi _{+}^{0}|^{2}\right) +V(\mathbf{r})-\mu _{+},\\
&\begin{aligned}
A_{12} &=g_{11}\,\left( \phi _{-}^{0}\right) ^{2}, & A_{34} &=g_{22}\,\left( \phi _{+}^{0}\right) ^{2},\\
A_{13} &=g_{12}\,\phi _{-}^{0}\left( \phi _{+}^{0}\right) ^{\ast }, & A_{14} &=g_{12}\,\phi _{-}^{0}\phi _{+}^{0}.
\end{aligned}
\end{align*}
Similar to~\cite{boulle2020deflation}, we decompose \cref{eig_prob} into real and imaginary parts to solve the discretized $8\times 8$ block matrix eigenvalue problem. We employ the same piecewise cubic finite element discretization as the one used for solving the NLS system and solve the resulting eigenvalue problem using a Krylov--Schur algorithm with a shift-and-invert spectral transformation implemented~\cite{stewart2002} in the Scalable Library for Eigenvalue Problem Computations (SLEPc)~\cite{hernandez2005slepc}. We then decompose the eigenfrequencies $\omega\in\mathbb{C}$ into real and imaginary parts as $\omega=\omega_r+i\omega_i$. A state is considered spectrally stable at the chemical potentials $(\mu_-,\mu_+)$ if the eigenfrequencies have imaginary parts satisfying $|\omega_i|<10^{-3}$.

To verify the spectral stability or weak instability of a state $\phi^0_{\pm}$ discovered by deflation, we integrate the time-dependent NLS system~\eqref{eq_time_eq} until $T=240$ by perturbing $\phi^0_{\pm}$ along its most unstable eigendirection similarly to \cref{lin_ansatz}. We then select $\epsilon=10^{-2}$ and use $\psi^{(0)}_\pm(\mathbf{r})\coloneqq \tilde{\Phi}_\pm(\mathbf{r},t=0)$ as initial state for the time-integration of the system. The time discretization of the system is performed using a modified Crank-Nicolson method~\cite{delfour1981finite} with a time-step $\Delta t = 10^{-2}$. As for the single component nonlinear Schr\"odinger equation~\cite{delfour1981finite}, one can show that this time-stepping scheme conserves the atomic number of each component of the state and its energy.

\section{Numerical Results}
\begin{figure}[htbp]
\center
\vspace{0.2cm}
\begin{overpic}[width=\figwidth]{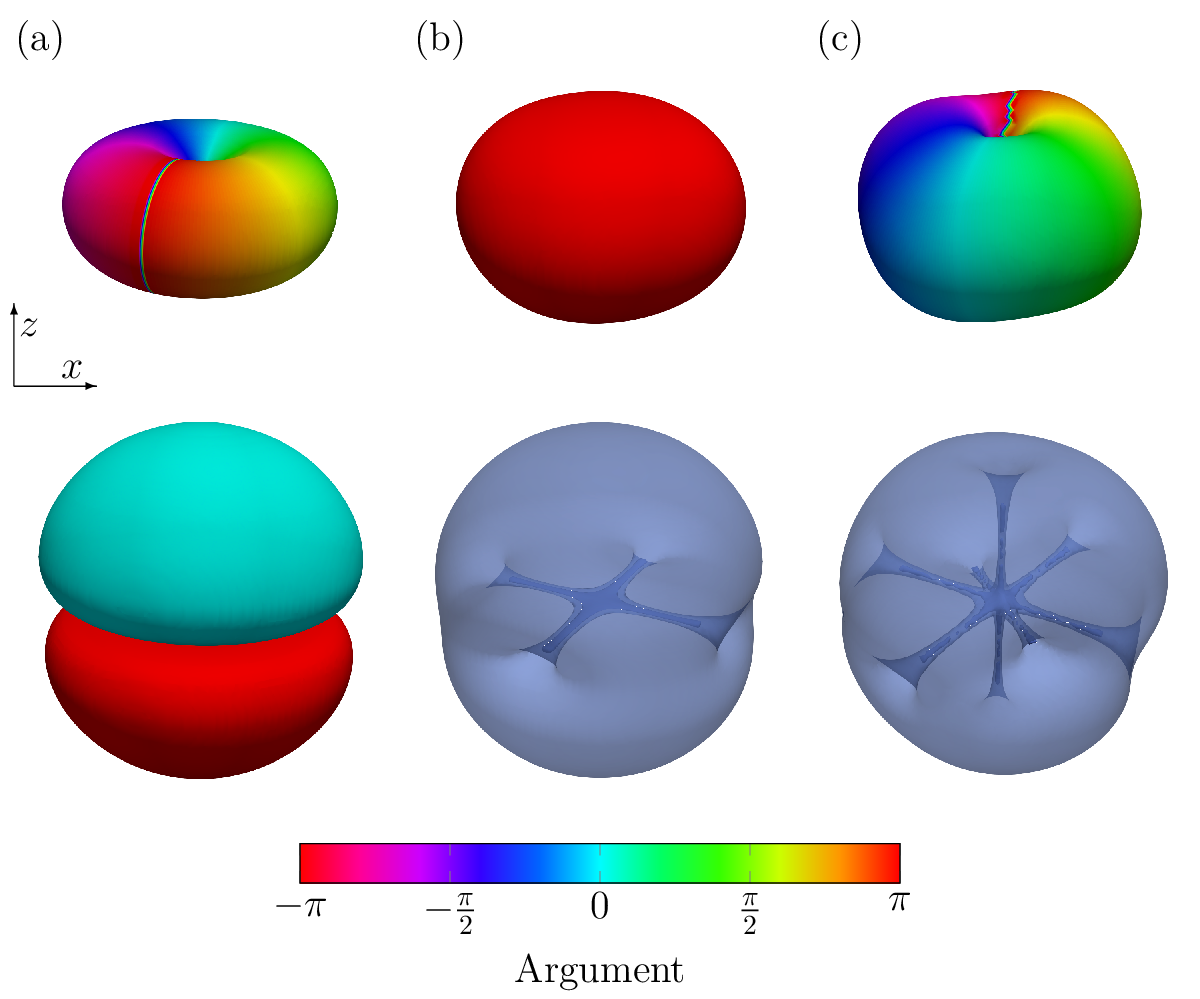}
\end{overpic}
\caption{Selection of three unstable steady-states whose individual components have been identified in previous works on the one-component GP equation. The top (resp.~bottom) row illustrates the $-$ (resp.~$+$) component of the solution. The colors represent the argument of the solution on the isosurface of magnitude $0.2$, \emph{i.e.}, $|\phi_-(x,y,z)|^2=0.2$ and $|\phi_+(x,y,z)|^2=0.2$. Whenever appropriate, we display the density isosurfaces at density $0.2$ with opacity $0.5$ to visualize the vortex structure of the component.}
\label{fig_simple_states}
\end{figure}

We now illustrate some solutions discovered with deflation. Our procedure leads to the discovery of 150 distinct solutions to \eqref{eq_GP} with complex structures. We then conduct a BdG stability analysis to focus on the most physically relevant ones and partition our findings into three broad categories. The first set of our results is shown in \cref{fig_simple_states} and captures a palette of unstable states partially identified in earlier works on one-component 3D systems. This illustrates that already some complex building blocks can be assembled into relevant stationary two-component solutions that the method can identify \emph{without prior knowledge} of associated theoretical or numerical constructions.

We recognize in \cref{fig_simple_states}(a) a dipole solitary wave in the second component coupled with a vortical pattern in the first component. It is helpful to utilize a Cartesian notation $|k,m,n\rangle$, highlighting the number of nodes that exist in each $x,y,z$ direction, to classify these states as a superposition of eigenstates near the linear limit. In this limit, this is the natural classification
equivalently representing the order of the polynomial modulating the Gaussian envelope of the states. In that notation, the first component consists (at low density, i.e.,~near the linear limit) of $|1,0,0\rangle + i|0,1,0\rangle$, while the second one represents $|0,0,1\rangle$. This state is unstable with a growth rate of $\omega_i\approx 0.81$. Similarly unstable ($\omega_i \approx 0.30$) in panel (b) is a so-called Chladni soliton~\cite{mateo2014chladni,mateo2015stability} of a one-species condensate in the second component, coupled to a ``ground state'' (a nodeless cloud) in the first component. Finally, panel (c) reveals a single vortex pattern similar to panel (a) in the first component, while the second is an example of a star pattern, reminiscent of the ones described in~\cite{boulle2020deflation,crasovan2004three}, with the following linear combination close to the linear limit: $|2,0,0\rangle - |0,2,0\rangle + i[|2, 0, 0\rangle - |0, 0, 2\rangle]$. This state has a growth rate of $\omega_i\approx 0.26$ and is unstable.

\begin{figure}[htbp]
\center
\vspace{0.2cm}
\begin{overpic}[width=\figwidth]{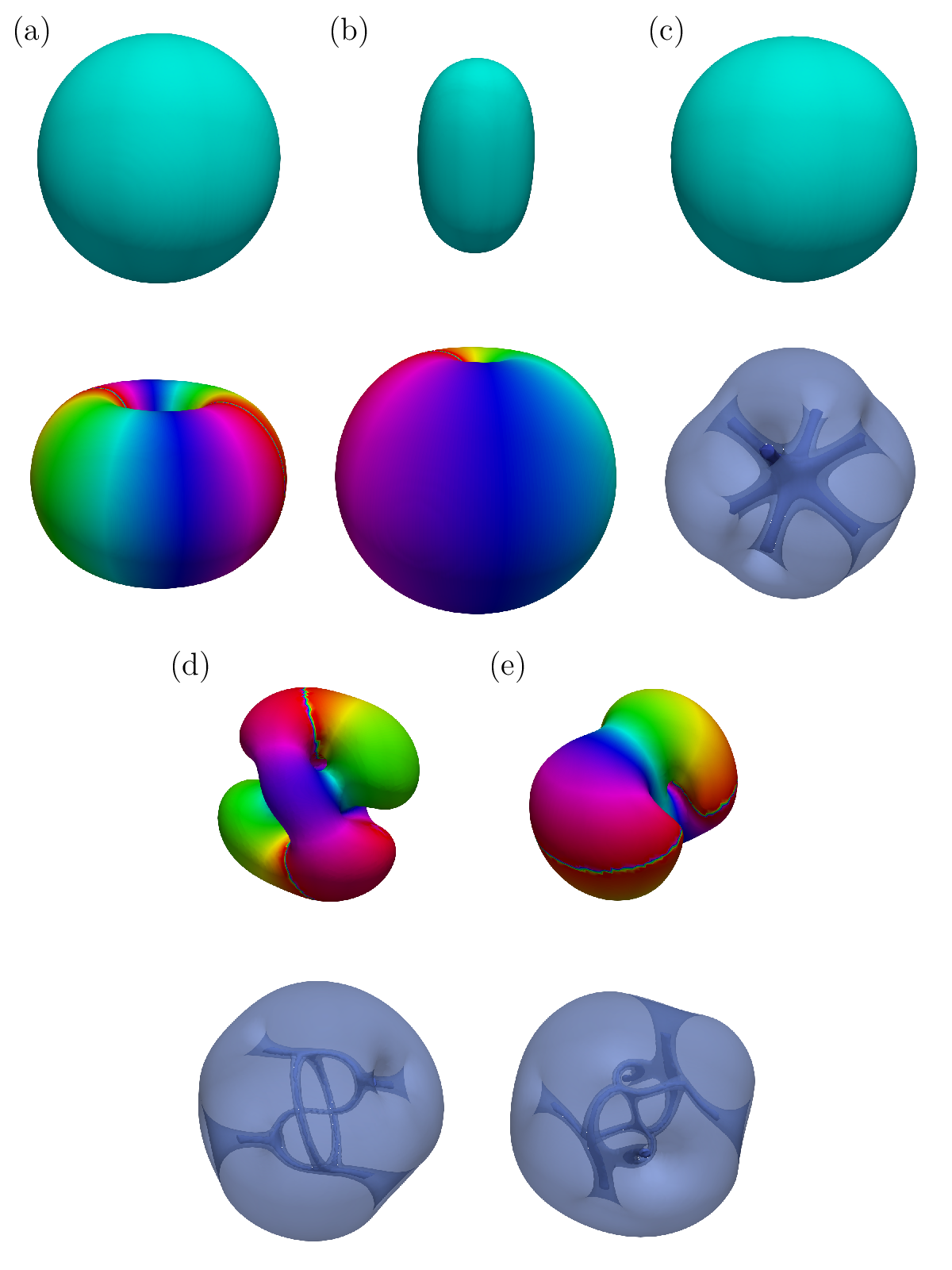}
\end{overpic}
\caption{Five stable states to the NLS system discovered by deflation.}
\label{fig_stable_state}
\end{figure}

\subsection{Dynamically stable states}
A more elaborate set of dynamically stable states has been identified and presented in \cref{fig_stable_state}. Here, we illustrate five distinct states identified as spectrally stable over {\it wide} parametric regimes. More specifically, the parameter regime we investigate is to linearly interpolate both chemical potentials, $\mu_-$ and $\mu_+$, towards their low-density limits at $\mu_{0-}$ and $\mu_{0+}$ (e.g.,~$(\mu_{0-},\mu_{0+})=(1.5,3.5)$ in \cref{fig_stable_state}(a)). Some of the resulting (widely) stable states are straightforward to interpret. For instance, \cref{fig_stable_state}(a) features a fundamental state in the first component that is complemented by a vortex of topological charge $l = 2$ in the second component. \cref{fig_stable_state}(b) shows a vortex of topological charge $l = 1$ in the second component harboring an effective bright soliton within a vortex-bright line configuration~\cite{wenlo}; such states have been explored as ``filled-core vortices'' already since the early experimental studies of~\cite{PhysRevLett.85.2857}. Then, in panel (c), we find a stable vortex star in the second component~\cite{crasovan2004three} coupled to a ground state of the first component.

\begin{figure}[htbp]
\center
\vspace{0.2cm}
\begin{overpic}[width=\figwidth]{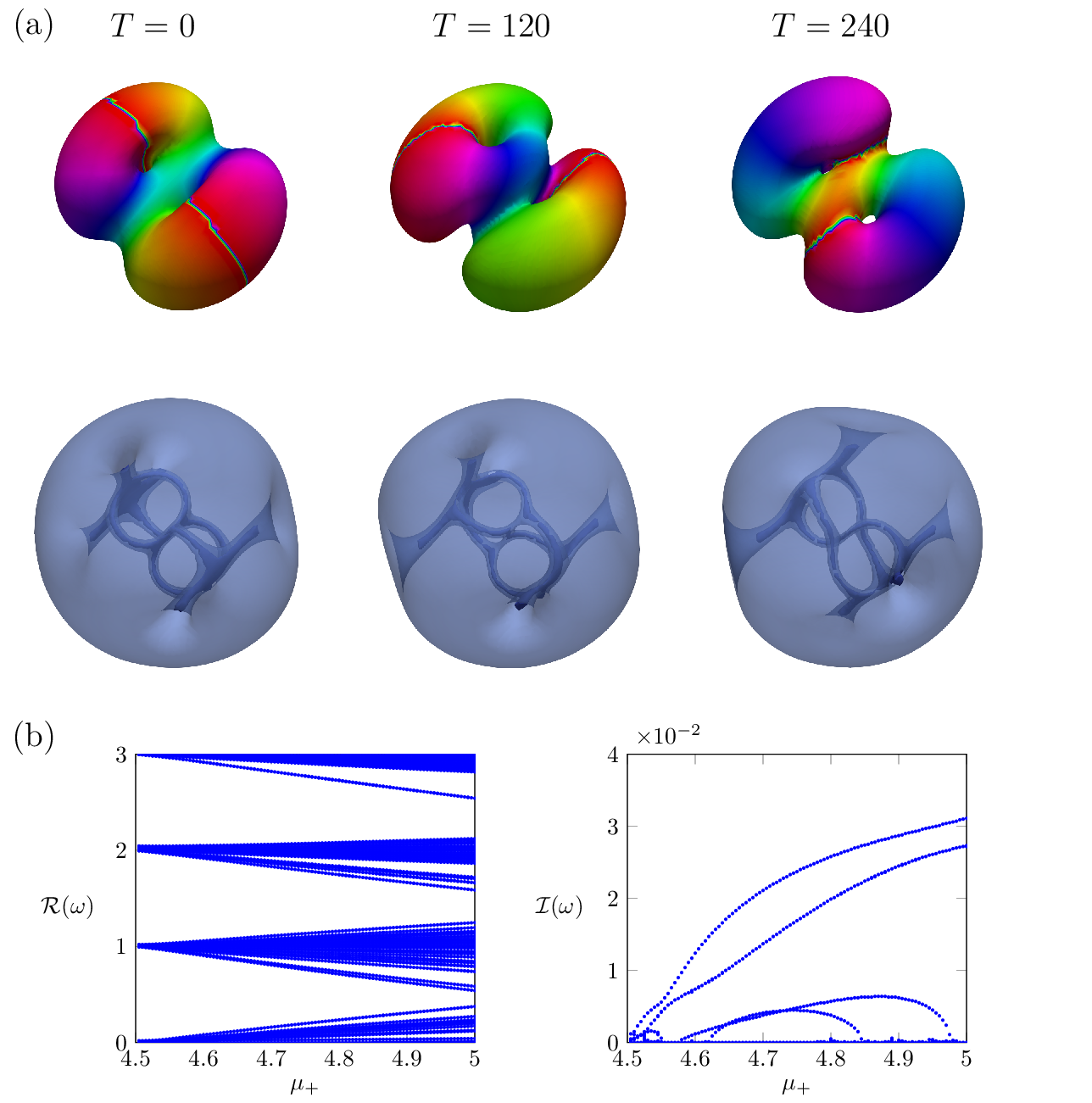}
\end{overpic}
\caption{(a) Snapshots of the time evolution of a weakly unstable state, initially perturbed along its most unstable eigendirection. The top and bottom rows respectively display the first and second components of the state. (b) Stability analysis of the steady state as the chemical potentials approaches $(3.5,4.5)$.}
\label{fig_129}
\end{figure}

\begin{figure*}[htbp]
\center
\begin{overpic}[width=\textwidth]{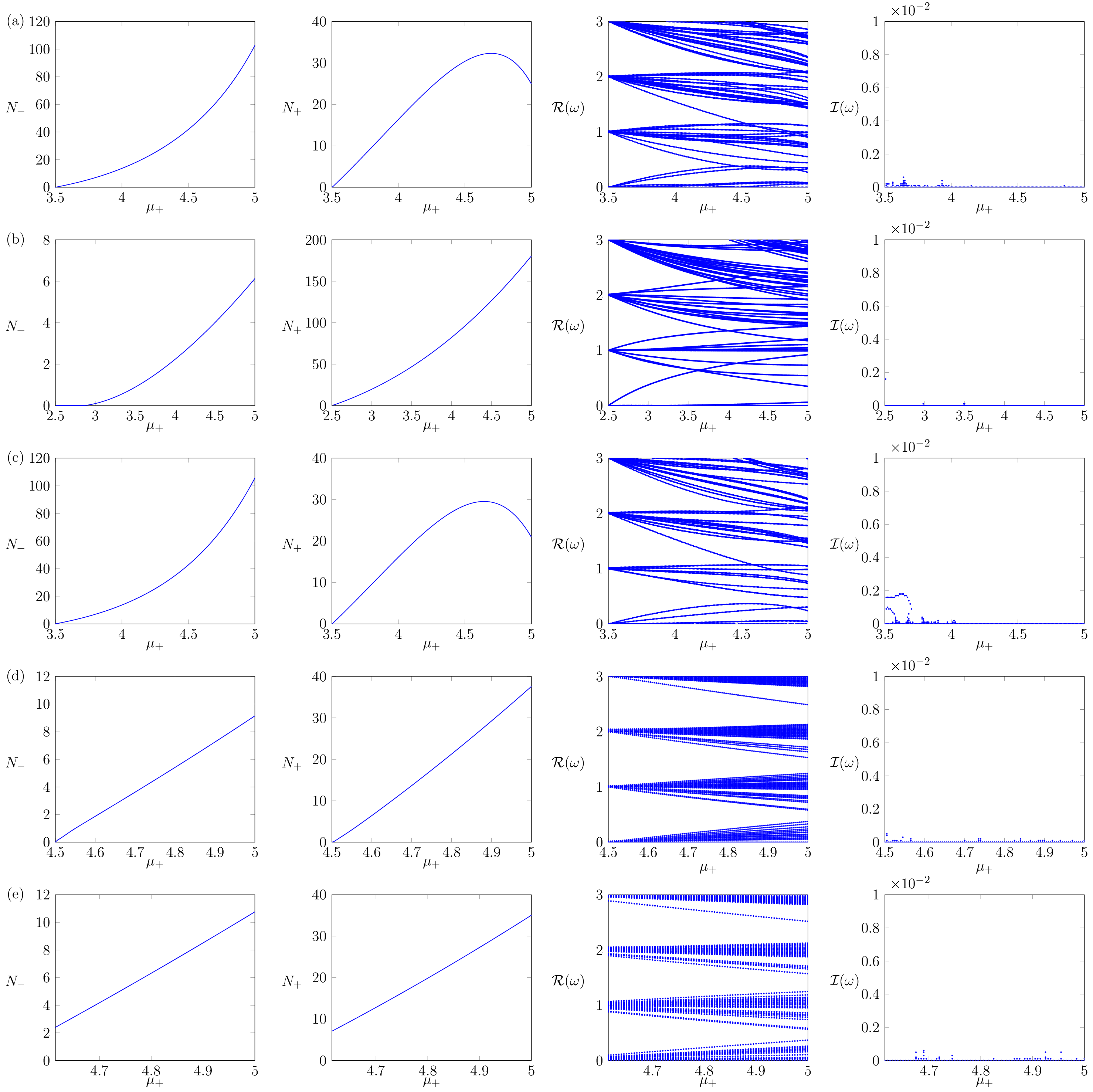}
\end{overpic}
\caption{Continuation and stability analysis of the states presented in \cref{fig_stable_state}. Each panel features the atomic numbers $(N_-,N_+)$ of the components of the state as well as the real and imaginary parts of the associated eigenfrequencies. The linear limit for the states presented in panels (a) to (e) are respectively $(\mu_-,\mu_+)\to (1.5,3.5)$, $(1.5,2.5)$, $(1.5,3.5)$, $(3.5,4.5)$, and $(3.5,4.5)$.}
\label{fig_stability}
\end{figure*}

While one can argue that the above states are perhaps ones that can be expected in the two-component realm based on our single-component experience, this is far from obvious in the context of panels (d) and (e) of \cref{fig_stable_state}. The structure of panel (d) contains, in turn, two vortex lines in the first component that are coupled to a second component featuring an S-shaped vortex attached to a vortex ring, as well as two additional U-shaped vortex lines. Interestingly, such a state in a single-component was also obtained in our previous work~\cite{boulle2020deflation} and was weakly unstable with a growth rate of $\omega_i\approx 5\times 10^{-2}$ at $\mu = 5$, illustrating that coupled systems may stabilize BEC configurations. \cref{fig_stable_state}(e) represents an especially complex, topologically charged configuration, where the vortex ring of the first component connects to an anti-symmetric pattern reminiscent of a pair of vortex-based slings, each held by a vortex line. Importantly, despite the elaborate multi-vortical structure of these two configurations, our computations identify them as a spectrally and dynamically stable. As an illustration, we report in \cref{fig_stability} the continuation of the atomic numbers $(N_-,N_+)$ for each of the two components of the state presented in \cref{fig_stable_state} as a function of the chemical potential $\mu_+$. The real part of the relevant eigenfrequencies in the third column of \cref{fig_stability} showcase the excitation frequencies, while the absence of imaginary eigenfrequencies (for most parameter values) in the fourth column indicates stability of the corresponding states within our BdG analysis.

\subsection{Weakly unstable states}
\begin{figure}[htbp]
\center
\vspace{0.2cm}
\begin{overpic}[width=\figwidth]{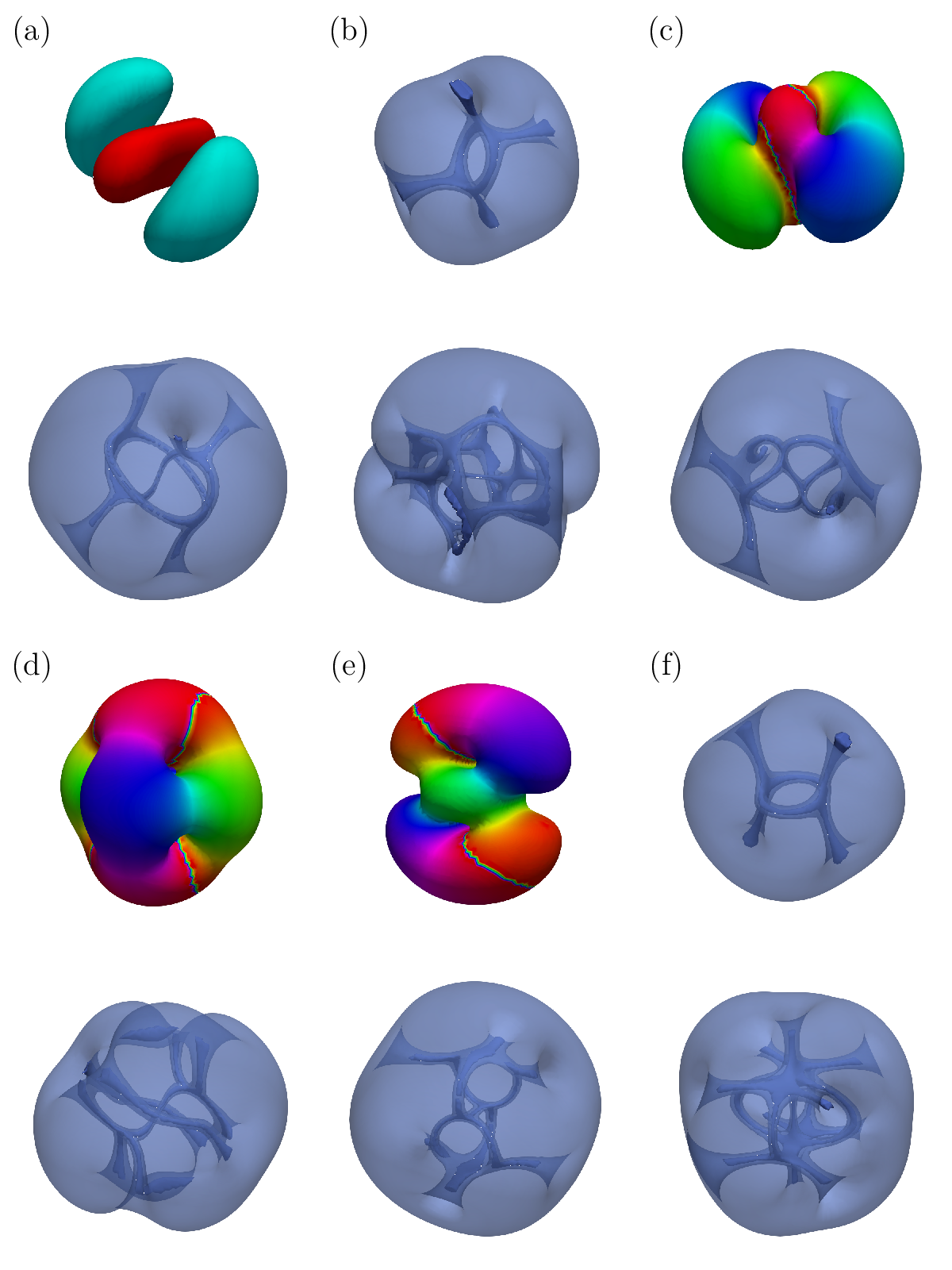}
\end{overpic}
\caption{Exotic weakly unstable solutions discovered by deflation which are within windows of experimental
observability with growth rate $\omega_r<6.7\times 10^{-2}$.}
\label{fig_weak_state}
\end{figure}

In addition to these stable states, our deflation search yields a considerable wealth of weakly unstable states such as the one shown in \cref{fig_129}(a) at time $T=0$. Here, we observe a pair of vortex lines in the first component, coupled to a labyrinthine network involving multiple vortex rings, as well as S-shaped and U-shaped vortices (see the bottom row). Remarkably given the complexity of the state, yet in line with the BdG stability analysis, our dynamical simulations~\cite{delfour1981finite}, involving hundreds of oscillation periods of the trap, identifies the state as only very weakly unstable, with a particularly small growth rate of $\omega_i\approx 3.1\times 10^{-2}$ (see \cref{fig_129}(b), and the movie in the Supplementary Material). This is only one of many such states that our detailed time-integration simulations, which conserve $N_{\pm}$ and the energy to machine precision, appear to preserve over time scales that would be relevant for experimental observability. Indeed, we present in \cref{fig_weak_state} a selection of exotic weakly unstable states whose vortex structures involve structures well beyond the complexity of simple vortex rings and lines and rather extend to labyrinthine patterns of connected vorticity isocontours. The structures are sustained over a long time period with growth rates bounded by $\omega_i<6.7\times 10^{-2}$ at chemical potential values $(\mu_-,\mu_+)=(4,5)$. Despite their complexity, all of these configurations are expected to be experimentally observable following our spectral stability analysis.

\section{Conclusions \& Future Work}
We have discovered a wealth of complex states to the two-component 3D GP equations. The deflation approach allowed us to retrieve a number of states that, to the best of our knowledge, had not been obtained previously, despite extensive efforts in this direction~\cite{KAWAGUCHI2012253,siambook,KEVREKIDIS2016140}. Even more importantly, several configurations involving combined vortical patterns, such as ones with S-shaped, U-shaped, and ring-shaped vortex patterns were found to be stable and, hence potentially accessible by state-of-the-art experimental techniques.

While a number of additional unstable states have been retrieved, current experimental techniques have made substantial progress towards ``sculpting'' a diverse host of initial conditions and the unstable or transient dynamics of such states may lead to intriguing pattern formation phenomena in multi-component condensates.

Our deflation technique enables the identification of complex (and possibly topologically-charged) patterns in a variety of nonlinear, elliptic partial differential equation problems of the Schr{\"o}dinger class. The results were proposed in the experimentally tractable platform of atomic Bose--Einstein condensates. However, other settings where the nonlinear Schr{\"o}dinger equation is relevant can be equally well applicable. In fact, the same numerical techniques could be broadly applicable to a variety of other problems, such as reaction-diffusion ones~\cite{Smoller}, among others.

The present work paves the way for numerous further possibilities. On the one hand, the theoretical understanding of such complex states, including from the linear limit of small amplitude, is a feature which has been explored in 1D and 2D settings~\cite{siambook}, but not systematically in the 3D case, to the best of our knowledge. This is due to the computational complexity of the problem. In the two-component setting of the present work, there is recent experimental   motivation~\cite{PhysRevLett.120.135301} to explore not only the case where $g_{12}>0$ as done herein, but also that of $g_{12}<0$ (but weak enough to avoid mean-field collapse). Furthermore, while we have restricted our attention to the two-component pseudo-spinor case in this work, a significant volume of experiments has been recently focusing on 3-component spinor settings~\cite{KAWAGUCHI2012253,RevModPhys.85.1191}. It would be particularly interesting to extend the ideas presented herein in the latter setting.

\acknowledgments{This work is supported by the EPSRC Centre For Doctoral Training in Industrially Focused Mathematical Modelling (EP/L015803/1) in collaboration with Simula Research Laboratory and an INI-Simons Postdoctoral Research Fellowship (N.B.) and a London Mathematical Society Undergraduate Research Bursary URB-2021-21 (I.N.). This material is based upon work supported by the UK Engineering and Physical Sciences Research Council under Grants EP/W026163/1 and EP/R029423/1 (P.E.F.), the US National Science Foundation under Grants No.~PHY-2110030 (P.G.K.), and a Leverhulme Trust Visiting Professorship grant No.~VP2-2018-007. The authors are particularly grateful to Prof.~E. Charalampidis for his insights and earlier collaboration.}

\bibliography{biblio}

\end{document}